# Nuclear Gamma Radiation Caused by a Muon at rest in $^{152}$Sm

*V.M.Abazov[a], V.A.Gordeev[b], K.Grama[c], S.A.Kutuzov[a], Do Khoang Kyong[e], M.Lewandowski[d], B.M.Sabirov[a], G.E.Solyakin[b]*

a — Lab. of Nuclear Problems, JINR, Dubna, Russia.
b — Nuclear Physics Institute, St.-Petersburg, Russia.
c — University of Bucharest, Bucharest, Romania.
d — Physics Institute of M.C. Sklodovskaja University, Lublin, Poland.
e — Vietnam National University, Hanoi, Vietnam

Investigation of the interaction of muons with complex nuclei yields important results on the properties of nuclei and on their electromagnetic interaction with muons happening to be in the Coulomb field of the nucleus [1, 2].

The Bohr orbit of a muon is ~207 times closer to the nucleus than that of an electron, the nucleus is transparent-like to it, and it spends a large part of its time inside the nucleus, especially in the case of medium and heavy nuclei. For this reason the muon serves as a unique instrument in studies of the charge distribution and of other properties of nuclei. The following are processes which yield information relevant to the properties of nuclei:

1. **Hyperfine interactions** that are responsible for dynamic effects of virtual excitation of the nucleus, due to the perturbative electromagnetic field of the muon, and result in corrections to the wave functions and energy levels of the muonic atom. They are essential for the lower states of the muonic atom, when the muon can spend a significant part of time inside the nucleus. Hyperfine interactions may be magnetic and electric quadrupole interactions. The magnetic interaction is due to interaction of the total angular momentum of the nucleus $F = I + J$ with the magnetic field created by the muon in the nucleus. The magnetic hyperfine structure constant for medium and heavy muonic nuclei is of the order of magnitude of several keV [3], which is significantly lower than corrections due to the "dynamic hyperfine splitting" caused by the muon-nucleus electric quadrupole interaction [4]. Enhancement of the deformation of the nucleus is accompanied by a decrease in the absolute excitation energy of the nucleus and an enhancement in its quadrupole angular momentum of and in the *E2* transition probabilities, which results in favourable conditions for resonance-like interaction between the muon and the nucleus. Investigation of dynamic hyperfine structure in the X-ray spectra of muonic atoms of deformed nuclei can yield information on the quadrupole angular momenta of excited nuclear states and on *E2* excitation probabilities.

If the excitation energy of a certain state of a nucleus exhibiting the same spin and parity as its ground state nearly coincides with its *2S→1S* transition energy, then dynamic monopole excitation of the nucleus will occur [5]. This effect occurs in nuclei within a limited range, since an increase in $A$ is accompanied by a decrease in the nuclear vibrational energy, while the $2S \rightarrow 1S$ transition energy increases. Dynamic $E0$ excitation has not yet been observed [6].

2. **Radiationless excitation of the nucleus** to the continuous spectrum was first predicted by D.F.Zaretsky [7] and, then, observed in Dubna [8]. Energy conservation allows this process only in the most heavy muonic atoms. If radiationless capture occurs, then nuclear decay via various channels is much more rapid, than nuclear capture of the muon. Fission represents one of the main transition channels to the continuous spectrum, and after fission the muon may be captured by one of the fission fragments and form a new muonic atom in an excited state [9]. Muonic X-ray emission from a fragment was recently observed in radiationless fission of the $^{238}U$ nucleus [10].

When the muon reaches the 1S state, its further destiny depends on two competing processes: 3. on **nuclear capture** and 4. on the μ- → e- + $υ_μ$ + $υ_e$ **decay.**

3. **Nuclear capture** of the muon is a weak semileptonic process. Intense studies of muon capture by complex nuclei aimed at testing weak interaction theory have revealed that issues related to nuclear structure are of great importance [2]. All aspects of muon capture in nuclei have been dealt with in detail and a thorough analysis of experimental data is presented in ref.[26].

4. **The decay of a bound muon** on the K-orbit of a mesonic atom is a process that competes with nuclear capture. If $τ_0$ is the lifetime of the muon on the K-orbit, then the rate of this process $λ_0 = 1/τ_0 = λ_c + Qλ_d$, where $λ_c$ is the nuclear capture rate and $λ_d$ is the decay rate, $Q$ is the Huff parameter [11]. The relationship between $λ_c$ and $λ_d$ changes drastically as $Z$ increases [12]: while in the case of small $Z$ muon decay is the dominant process, at $Z=11$(Na) the two processes become equal, and when Z~25—30 about 90% of the muons are captured. The Huff parameter $Q$ determines the decrease in the decay rate of a bound muon with respect to the decay of a free muon and depends on three factors: the ¯system (i.e. in the centre-of-mass system of the nucleus), and Coulomb interaction of the muon and electron with the nucleus. The parameter $Q$ is particularly significant in the case of heavy nuclei.

5. **The decay of a muon on the K-orbit in interaction with the nucleus** can be interpreted as a sharp switchoff of the spherical Coulomb field created around the nucleus by the muon on the orbit of angular momentum $ℓ = 0$ resulting in an electric shakeup of the nucleus. The energy released in this process significantly exceeds the energies of low-lying nuclear excitations in medium and heavy deformed nuclei. A feature peculiar to this process is that monopole excitation of the nucleus should dominate. This phenomenon was first noticed by I.S.Batkin [13]. He calculated the excitation probabilities for the first $0^+$ states in several rare-earth nuclei applying the Davydov-Chaban droplet model and obtained ω(E0)~$10^{-2}$ per decay event. Calculations performed by I.A.Mitropolsky utilizing the microscopic model of the nucleus yielded results by 1-2 orders of magnitude lower [14]. The upper limit obtained in the first experiment in search of monopole excitation in the $^{152}Sm$ yielded the following value for the excitation probability of the $0^+$ state: $w(E0) < 5·10^{-3}$ [15].

The present work continues the investigation of the γ-radiation of a $^{152}Sm$ nucleus, in which a muon has stopped. The choice of the $^{152}Sm$ nucleus as a target is due to the following:

a) this is a rare-earth transition nucleus with noticeable deformation, and, consequently, the energy of the first excited $0_2^+$ state is low (685 keV). The transition energy is present in the denominator of the corresponding matrix element [14];

b) the ratio between the probability of conversion de-excitation of the $0^+$ level, $w_e$, and the probability of radiation transition, $\omega_\gamma$, is 0.02 [15].

The measurement was performed in the low-background laboratory of the JINR Laboratory of Nuclear Problems. The beam intensity was ~4·10$^4$μ/sec, the number of muon stops in the target weighing 117g amounted to (7÷10)·10$^3$ sec$^{-1}$. A powder of samarium trioxide $Sm_2O_3$ enriched up to 98% was used.

The target was made in the form of an upside-down letter $V$ permitting enhancement of the thickness of the material in the path of the muons and decreasing the path for γ-quanta (the $Ge(Li)$ detector was situated under the target). The sketch of the layout of MEGA experimental installation is shown in Fig.1. Its construction and operation principle are described in refs.[15] and [16] for extracting gamma and muonic X-rays.

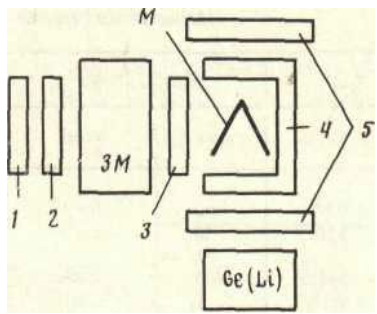

**Fig.1**. *Schematic Layout of MEGA installation $C_1$-C4, $C_6$ - plastic counters; C5 - Cherenkov counter, 3M - muon beam degrader, M – $Sm_2O_3$ target, Ge(Li) – semiconductor detector.*

Energy calibration of the spectrometric system involving a $Ge(Li)$ detector with a sensitive volume ~50cm$^3$ permitted determination of the energy of statistically reliable lines with a precision of 2-3 tens of eV. Fully calibrated $^{182}Ta$, $^{152}Eu$, $^{192}Ir$, $^{133}Ba$, $^{110}Ag$, and $^{56}Co$ γ- sources were used for calibration with respect to absolute efficiency. For eliminating corrections due to the complicated geometry of the extended target each source was placed successively in 12 zones into which the planes of the support for the target were divided. After calculating the absolute intensity for each isotope taking into account its certification time, its halflife and the target irradiation time, all the obtained points for all the sources were approximated by the following logarithmic four-parameter function [17]:

$$U = a_1 + a_2 \ln t + a_3 t + a_4 \Theta(t-1) t \ln t,$$

where t= $E/E_0 (E_0 = m_e c^2)$. The calibration points scaled to the respective calculated absolute efficiencies are shown in Fig.2 for the utilized γ-sources. The measurement errors were within the sizes of the points.

Unlike the plastic counter used for the detection of electrons and described in ref.[15], a Cherenkov counter utilizing *TF-1* lead glass and a FEU-143 photomultiplier operating in the single-electron mode were utilized in the present experiment. The *TF-1* refraction coefficient was $n_D$ = 1.65, i.e. the amount of photons emitted by a *β* = 0.99 electron in the 40nm÷70nm range was $N_{phot}$≈330 phot/cm [23].

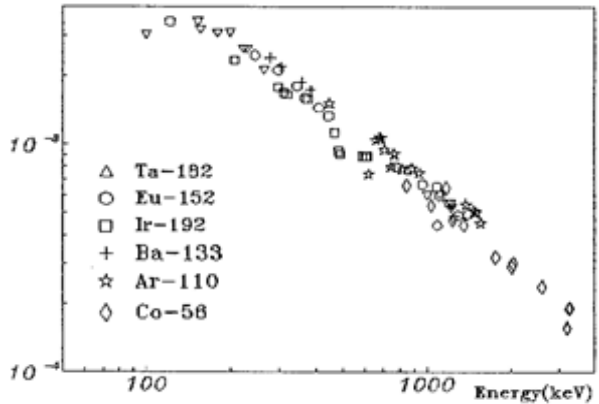

**Fig.2**. *Absolute efficiency of the Ge(Li) detector versus the ɣ energy.*

The transmission coefficient of *TF-1* for visible light ~0.9 [18]. Therefore, if our counter is *5cm* thick, its efficiency can be considered to be in the vicinity of 0. The detection threshold for the *ɣ*-spectra was set at a level of 85-90 keV. The measured mesonic *X*-ray spectra coincide with data presented in the literature [19]. Selection of the gamma-spectrum related to electrons was performed by coincidence of a time signal from the *Ge(Li)* detector inside a 1 *μsec* gate initiated by the stop of a muon and the signal from the Cherenkov counter, also inside the gate from the μ- stop [16]. The two-dimensional time distribution of ɣ — e coincidences is shown in Fig.3-4.

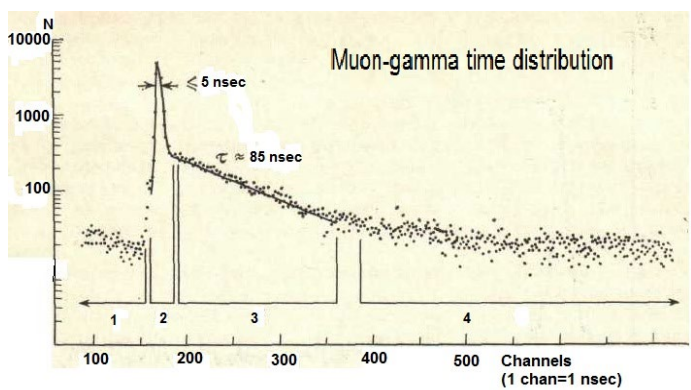

**Fig.3**. *Muon-gamma time distribution, measured in 1+2+3-4+tɣ coincidence.*

An enhanced concentration of events is to be noted in the lower right-hand part of the distribution. These are ɣ —e events corresponding to the time after departure of the electron. One tends to think of possible interaction in the final state. But this is difficult to imagine, since the propagation time of the electron in the space occupied by a nucleus of radius *R* is $_1τ$~ *R/c*~$10^{-23}$ sec, while the characteristic time of nuclear oscillations $\tau_2 \sim \hbar/E \sim 10^{-21}$ sec, i.e. $\tau_1 \ll \tau_2$ [13]. But the energy spectra corresponding to the lower part to the right of "trail" and to the upper part to the left of the "trail", respectively, somewhat differ. Therefore this phenomenon deserves special investigation. The frame indicating an exponential segment along the "trail" of $\gamma - e$ coincidences corresponds to the lifetime of a muon on the K-orbit of the $^{152}$Sm muonic atom. It is essential to find the optimum position of the starting point of the time window for delayed events to make the contribution to these events from prompt events having a Gaussian distribution as small as possible, so as to reduce to a minimum the loss due to the area of the exponential.

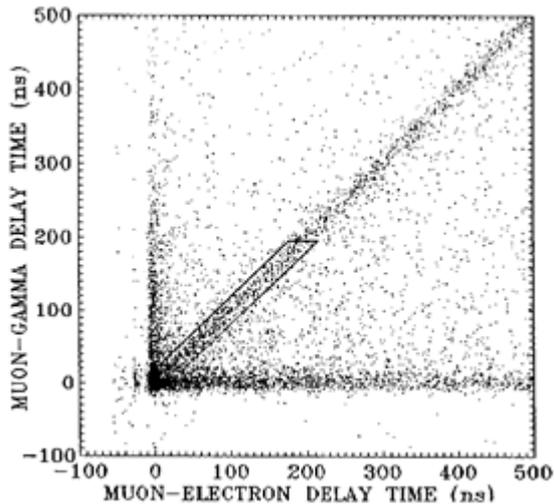

By appropriate selection such a position was found, for which the area under the exponential in this window amounted ~78% of the total area, Fig.4

**Fig.4**. Spectrum of $\gamma - e$ coincidences

The main goal of the search was the $E2$ $\gamma$-transition ($E_\gamma = 563.2$ keV) between the $0_2^+$ and $2_1^+$ levels in coincidence with the decay electron. A segment of such a spectrum in the ~500-600 keV range is shown in Fig.5. The arrow indicates the position of the 563 keV peak. The average background at this point is $N_{bgrd} \approx 40$.

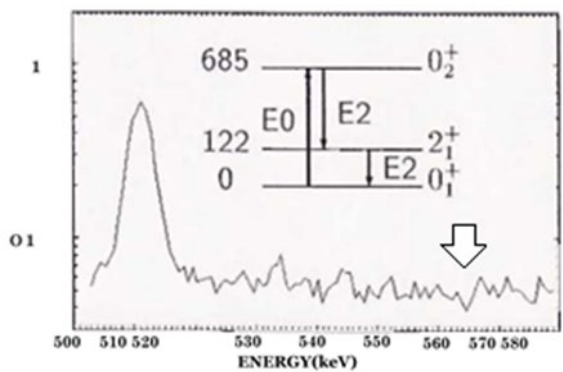

**Fig.5**. $\gamma$ - spectrum in the energy region of the $0_2^+ \rightarrow 2_1^+$ transition for events indicated by the frame in Fig.3

The counting rate expected at the 563 keV peak is determined by the following expression:

$$N = N_\mu \cdot \omega_{ac} \cdot a \cdot \omega_d \cdot \omega(0^+) \cdot B \cdot \Omega_e \cdot \varepsilon_e \cdot \varepsilon_\gamma \cdot T_\gamma \cdot A,$$

where $N_\mu$ — is the whole number of $\mu$—stops, $\approx 2 \cdot 10^9$;
$\omega_{ac}$ — the probability of atomic capture by $Sm$, $=0.76$ [22];
$a$ — the isotopic enrichment with $^{152}Sm$, $=0.98$;
$w_d$ — the decay probability of a muon on the K-orbit of $^{152}Sm$, $\approx 3.83 \cdot 10^{-2}$;

$w(0^+)$ — the excitation probability of the $0^+$ level in $^{152}$Sm per decay event, $3 \cdot 10^{-4}$;
$B$ — the $0^+$ de-excitation branching via the $E2$ γ-transition, ≈0.98;
$\Omega_e$ — the solid angle covered by the electron counter, ≈0.42;
$\varepsilon_e$ — the electron detection efficiency, ≈0.9;
$\varepsilon_\gamma$ — the absolute efficiency of the γ-spectrometer; $\varepsilon_\gamma \approx 10^{-3}$ for $E_\gamma$=563keV;
$T_7$ — a coefficient taking into account the overlapping of the exponential and the Gaussian distribution, ~0.78;
$A$ — the absorption coefficient of the target for γ-quanta with $E_\gamma$=563 keV, ≈0.86.

Thus, $N \approx 4.7$, which at a 90% confidence level yields for the experimental quantity $w(E0)$ an upper limit $< 1.5 \cdot 10^{-3}$. This is 3.3 times lower than our previous result ($< 5 \cdot 10^{-3}$), but still does not reach the theoretical estimation ($\approx 3 \cdot 10^{-4}$).

Above it was mentioned that dynamic $E2$ muon-nucleus interaction inside the nucleus may result in excitation of low-lying rotational levels. The probability for the $^{152}$Sm nucleus to remain in the $2_1^+$ state after completion of the muonic cascade is 0.30 [20]. The de-excitation time of the $2_1^+$ nuclear level ($E_\gamma$ = 121.78 keV) is about =$1.44 \cdot 10^{-9}$ sec, the lifetime of the muon on the $K$-orbit is $\approx 8 \cdot 10^{-8}$ sec. This means that the $2_1^+ \to 0_1^+$ nuclear transition takes place when the muon is present on the $K$-orbit of the muonic atom. The perturbation introduced by the muon alters the radius of the excited nucleus. The change occurring in the nuclear potential has been termed the "isomeric shift". Experimentally observed isomeric shifts in various nuclei have been reported in refs.[19a,20-22]. The results are spread out in the range from 350eV up to ≥1keV. Measurements of the isomeric shift in the $2_1^+$ level of $^{152}$Sm have yielded the following values: $\Delta E_{is}$ = 1.03±0.15keV, 560±60eV and 550±70eV [20,21]. Our measurement gave $\Delta E_{is}$ = 820±40eV.

## Conclusion

Thus, in the present work, a new upper limit has been obtained for the probability of monopole excitation of the $^{152}$Sm nucleus due to the decay of a bound muon on the K-orbit of a muonic atom: $w(E0) < 1.5 \cdot 10^{-3}$. The isomeric shift for the $2_1^+$ level (121.78 keV) has been measured to be $\Delta E_{is}$ = 820±40eV (preliminary result) [25].

This experiment should evidently be repeated with the apparatus having been significantly upgraded so as to permit high-statistics measurements in conditions of a very low background. Information on the phenomenon investigated in the present work is of interest from the point of view of analysis of neutrinoless μ→e conversion in Mu2e and COMET experiments.

The authors are grateful to the Direction of the JINR Laboratory of Nuclear Problems for support of this work. We wish to thank Dr. I.A. Mitropolsky for constant interest in our work and fruitful discussions at all the stages of the experiment.

*Contact:* sabirov@jinr.ru